\documentclass[%
reprint,
 amsmath,amssymb,
 aps,
]{revtex4-1}
\usepackage{graphicx}
\usepackage{braket}
\usepackage{bm}
\usepackage{amsmath}
\begin{document}
\title{Three-body decay of linear-chain states in $^{14}$C}
\author{T. Baba$^1$ and M. Kimura$^{1,2}$}
\affiliation{$^1$Department of Physics, Hokkaido University, 060-0810 Sapporo, Japan\\
$^2$Reaction Nuclear Data Centre, Faculty of Science, Hokkaido University, 060-0810 Sapporo, Japan}
\date{\today}

\begin{abstract}
 The decay properties of the linear-chain states in $^{14}$C are investigated by using the
 antisymmetrized molecular dynamics. 
 The calculation predicts two rotational bands with linear-chain configurations having
 the $\pi$-bond and $\sigma$-bond valence neutrons.
 For the $\pi$-bond linear-chain, the calculated excitation energies and the widths of
 $\alpha$-decay to the ground state of $^{10}{\rm Be}$ reasonably agree with the
 experimental candidates observed by the $\alpha+{}^{10}{\rm Be}$ resonant scattering.
 On the other hand, the $\sigma$-bond linear-chain is the candidate of the
 higher-lying resonant states reported by the break-up reaction.
 As the evidence of the $\sigma$-bond linear-chain,
 we discuss its decay pattern.
 It is found that the $\sigma$-bond linear-chain not only decays to the excited band of
 $^{10}{\rm Be}$ but also decays to the three-body channel of
 $^{6}{\rm He}+\alpha+\alpha$, and the branching ratio of these decays are
 comparable.
 Hence, we suggest that this characteristic decay pattern is a strong signature of the
 linear-chain formation and a key observable to distinguish two different linear-chains.   
\end{abstract}

\maketitle
\section{introduction}
Since the linear-chain configuration of 3$\alpha$ clusters (linearly aligned 3$\alpha$ clusters)
was suggested in 1950's \cite{mori56}, many studies have been devoted to search it in the excited states of $^{12}$C \cite{uega77,kami81,desc87,enyo97,tohs01,funa03,neff07,funa15,funa16}.
Nevertheless, no positive evidence has been obtained, and it is considered that the linear-chain
is unstable against the bending motion \cite{enyo97,neff07}.

Instead of $^{12}$C, neutron-rich C isotopes have attracted much interest in these decades as
new candidates of the linear-chain, because it is expected that the valence neutrons will play a
glue-like role and stabilize the linear-chain against the bending motion \cite{itag01}.
Theoretical studies predict the rotational bands with the linear-chain configurations in C isotopes \cite{oert03,oert04,itag06,suha10,maru10,furu11,suha11,baba14,zhao15,kimu16,baba16}.
It was pointed out that the motion of the valence neutron can be qualitatively interpreted in terms of the molecular orbits analogous to the Be isotopes \cite{seya81,oert96,itag00,yeny99,amd1,amd2}, which are called $\pi$- and $\sigma$-orbits.
Concurrently, rather promising candidates of the linear-chain were reported in $^{14}$C and $^{16}$C by several experiments \cite{gree02,bohl03,ashw04,pric07,free14,frit16,dell16}.

In our previous study \cite{baba16}, based on the antisymmetrized molecular dynamics (AMD) calculation, we pointed out that two positive-parity rotational bands in $^{14}$C have the linear-chain configurations.
The first band which we call $\pi$-bond linear chain has two valence neutrons in $\pi$-orbit and is built on the $0^+$ state at 14.6 MeV which is just above the $\alpha+{}^{10}{\rm Be}$ threshold but below the $2\alpha+{}^6{\rm He}$ threshold.
It was found that the energies and widths of the resonances observed by the $\alpha+{}^{10}{\rm Be}$ elastic scattering \cite{free14,frit16} qualitatively agree with the theoretical calculations \cite{suha10,suha11} including ours.
Therefore, they are regarded as the $\pi$-bond linear-chain candidates, although the experimental spin-parity assignment was somewhat ambiguous.
The other band named $\sigma$-bond linear chain has valence neutrons in $\sigma$-orbit and built on the $0^+$ state above the $\alpha+{}^{10}{\rm Be}$ and $2\alpha+{}^{6}{\rm He}$ thresholds.
It has more elongated linear-chain configuration and larger moment of inertia than the former band, but the experimental counterpart was not known at that time.

Quite recently, very interesting data were reported by two groups.
Yamaguchi {\it et al.} \cite{yama17} reported the result of the $\alpha+{}^{10}{\rm Be}$ elastic scattering and updated the information about the candidates of the $\pi$-bond linear chain.
The reported resonances look essentially same with those found in previous experiments \cite{free14,frit16}.
However, owing to the better statistics and larger angular coverage, they provided more reliable spin-parity assignment and evaluation of the decay widths.
The other experiment was reported by Tian {\it et al.} \cite{tian16} and Li {\it et al.} \cite{li17} who observed the resonances populated by $^9{\rm Be}({}^9{\rm Be},\alpha+{}^{10}{\rm Be})\alpha$ reaction.
In addition to the same resonances reported by Yamaguchi {\it et al.}, they found new resonances located {\it above} the $\alpha+{}^{10}{\rm Be}$ and $2\alpha+{}^{6}{\rm He}$ thresholds.
Based on the observed energies and decay pattern, these new resonances were suggested as the candidates of the $\sigma$-bond linear chain.

These new data motivated us to perform additional analysis and to summarize the calculated and observed properties of the linear-chain bands in $^{14}$C.
We investigated several decay modes of the linear-chain bands whose wave functions are obtained in our previous work \cite{baba16}.
By the comparison with the new data, it is found that the agreement between the calculated and observed $\pi$-bond linear-chain band is plausible.
It is also shown that the observed unique decay pattern of the resonances reported by Tian {\it et al.} agrees with the $\sigma$-bond linear-chain band, which is the first evidence for the existence of two different linear-chain bands with $\pi$- and $\sigma$-bonding.
In addition to these analysis, it was found that the $\sigma$-bond linear chain decays to the $^6{\rm He}+{}^{8}{\rm Be}$ channel as well as the $\alpha+{}^{10}{\rm Be}$ channel, and their branching ratios are comparable.
Hence, we suggest that the sequential three-body decay of $^{14}{\rm C}^*\rightarrow{}^6{\rm He}+{}^8{\rm Be}\rightarrow{}^6{\rm He}+\alpha+\alpha$ is an important evidence of the $\sigma$-bond linear chain.

The paper is organized as follows.
The AMD framework and the method to estimate the reduced widths amplitude for the $\alpha+{}^{10}{\rm Be}$ and $^6{\rm He}+{}^{8}{\rm Be}$ decays are explained in the next section.
In Sec. III, the excitation energies and decay widths of the linear-chain states are shown and compared with the observed data to suggest the assignment of the linear-chain bands.
In the last section, we summarize this work.

\section{theoretical framework}
In this study, we analyze the wave functions of linear-chain states obtained in our previous work \cite{baba16}.
For the sake of the self-containedness, in Sec. II. A., we briefly explain how those wave functions were calculated.
In Sec. II. B. and C., we explain the method to evaluate the decay modes of linear-chain states used in the present study.

\subsection{variational calculation and generator coordinate method}
We use the microscopic $A$-body Hamiltonian,
\begin{align}
 \hat{H} = \sum_{i=1}^A \hat{t}_i - \hat{t}_{c.m.} + \sum_{i<j}^A \hat{v}^N_{ij} + \sum_{i<j}^Z \hat{v}^C_{ij},
\end{align}
where the Gogny D1S interaction \cite{gogn91} is used as an effective nucleon-nucleon interaction $\hat{v}^N$.
The Coulomb interaction $\hat{v}^C$ is approximated by a sum of seven Gaussians. 
The kinetic energy of the center-of-mass $\hat{t}_{c.m.}$ is exactly removed. 

The variational wave function $\Phi^{\Pi}$ is a parity projected intrinsic wave function $\Phi_{int}$, and $\Phi_{int}$ is represented by a Slater determinant of single particle wave packets, 
\begin{align}
 \Phi^\Pi &= \hat{P}^\Pi\Phi_{int}, \\
 \Phi_{int} &= {\mathcal A} \{\varphi_1,\varphi_2,...,\varphi_A \},
  \label{EQ_INTRINSIC_WF}  
\end{align}
where $\hat{P}^\Pi$ denotes parity projector.
In this study, we focus on the positive-parity states ($\pi=+$).
$\varphi_i$ is the single particle wave packet which is a direct product of the deformed
Gaussian spatial part \cite{kimu04}, spin ($\chi_i$) and isospin ($\xi_i$) parts,  
\begin{align}
 \varphi_i({\bm r}) &= \phi_i({\bm r})\otimes \chi_i\otimes \xi_i, \label{eq:singlewf}\\
 \phi_i({\bm r}) &= \exp\biggl\{-\sum_{\sigma=x,y,z}\nu_\sigma\Bigl(r_\sigma -\frac{Z_{i\sigma}}{\sqrt{\nu_\sigma}}\Bigr)^2\biggr\}, \\
 \chi_i &= a_i\chi_\uparrow + b_i\chi_\downarrow,\quad
 \xi_i = {\rm p} \quad {\rm or} \quad {\rm n}.\nonumber
\end{align}
The centroids of the Gaussian wave packets $\bm Z_i$, the direction of nucleon spin $a_i, b_i$,
and the width parameter of the deformed Gaussian $\nu_\sigma$ are the variational parameters.  
The variational parameters are determined so that $E^\Pi$ which is a sum of the energy and constraint potential is minimized.
\begin{align}
 E^\Pi = \frac{\braket{\Phi^\Pi|\hat{H}|\Phi^\Pi}}{\braket{\Phi^\Pi|\Phi^\Pi}} 
 + v_\beta(\braket{\beta} - \beta)^2  + v_\gamma (\braket{\gamma} - \gamma)^2,
\end{align}
where $\braket{\beta}$ and $\braket{\gamma}$ are the quadrupole deformation parameters of the intrinsic wave function defined in Ref. \cite{kimu12}, and $v_\beta, v_\gamma$ are chosen to be sufficiently large value.
$E^\Pi$ is minimized by the frictional cooling method, and we obtain the optimized wave function 
$\Phi^\Pi(\beta,\gamma)=\hat{P}^\Pi\Phi_{int}(\beta,\gamma)$ 
which has the minimum energy for each set of $\beta$ and $\gamma$.

After the variational calculation, the eigenstate of the total angular momentum $J$ is projected
out,
\begin{align}
 \Phi^{J\Pi}_{MK}(\beta,\gamma) &=  \hat{P}^{J}_{MK}\Phi^\Pi(\beta,\gamma). 
\end{align} 
Here, $\hat{P}^{J}_{MK}$ is the angular momentum projector. 
Then, we perform the GCM calculation by employing the quadrupole deformation parameters $\beta$
and $\gamma$ as the generator coordinate.  The wave function of GCM reads,

\begin{align}
 \Psi^{J\Pi}_{Mn} = \sum_i\sum_Kc^{J\Pi}_{Kin}\Phi^{J\Pi}_{MK}(\beta_i,\gamma_i),\label{eq:gcmwf}
\end{align}
where the coefficients $c^{J\Pi}_{Kin}$ and eigenenergies $E^{J\Pi}_n$ are obtained by solving the
Hill-Wheeler equation \cite{hill54}.

\subsection{reduced width amplitude and decay width}
Using the GCM wave function, we estimate the reduced width amplitudes (RWA)
$y_{lj^{\pi}_n}(r)$ for the $\alpha+{}^{10}{\rm Be}$ and $^6{\rm He}+{}^{8}{\rm Be}$ decays which are defined as,
\begin{align}
 y_{lj^{\pi}_n}(r) = \sqrt{\frac{A!}{A_{\rm He}!A_{\rm Be}!}}
 \langle \phi_{\rm He}[\phi_{\rm Be}(j^{\pi}_n)Y_{l0}({\hat r})]_{J^\Pi M}
 |\Psi^{J\Pi}_{Mn}\rangle,\label{eq:rwa}
\end{align}
where $\phi_{\rm He}$ denotes the ground state wave function for $^4{\rm He}$ or $^6{\rm He}$, and $\phi_{\rm Be}(j^{\pi}_n)$ denotes the wave functions for daughter nucleus $^{10}{\rm Be}$ or $^{8}{\rm Be}$ with spin-parity $j^{\pi}_n$.
$Y_{l0}({\hat r})$ is the orbital angular momentum of the inter-cluster motion, and it is coupled with the angular momentum of Be$(j^{\pi}_n)$ to yield the total spin-parity $J^\Pi$.
$A_{\rm He}$ and $A_{\rm Be}$ are the mass numbers of He and Be, respectively.
The reduced width $\gamma_{lj^{\pi}_n}$ is given by the square of the RWA,
\begin{align}
 \gamma^2_{lj^{\pi}_n}(a) = \frac{\hbar^2}{2\mu a}[ay_{lj^{\pi}_n}(a)]^2,
\end{align}
and the partial decay width is a product of the reduced width and the penetration factor $P_l(a)$,
\begin{align}
 \Gamma_{lj^{\pi}_n} &= 2P_l(a)\gamma^2_{lj^{\pi}_n}(a), \quad
 P_l(a) = \frac{ka}{F^2_l(ka)+G^2_l(ka)}, 
\end{align}
where $a$ denote the channel radius, and $P_l$ is given by the Coulomb
regular and irregular wave functions $F_l$ and $G_l$. The wave number $k$ is determined by the decay
$Q$-value and the reduced mass $\mu$ as $k=\sqrt{2\mu E_Q}$.

\begin{figure}[h]
 \centering
 \includegraphics[width=0.8\hsize]{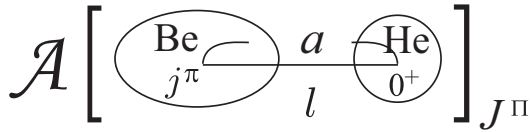}
 \caption{The schematic figure showing the Brink-Bloch wave function.} 
 \label{fig:illust_bbwf}
\end{figure}

To reduce the computational cost, we employ an approximate method given in
Ref. \cite{enyoRWA} to calculate Eq.(\ref{eq:rwa}).
In this method the antisymmetrization effect is neglected by choosing sufficiently large inter-cluster distance $a$, and RWA is approximated by the overlap between the GCM wave function and the Brink-Bloch wave function $\Phi_{BB}^{J\pi jl}(a)$ in which He and Be clusters are placed with the inter-cluster distance $a$ as illustrated in Fig. \ref{fig:illust_bbwf},
\begin{align}
 |ay_{lj^{\pi'}}(a)|^2 &\simeq \sqrt{\frac{\gamma}{2\pi}}
 |\braket{\Phi^{J\pi jl}_{BB}(a)| \Psi^{J\pi}_{Mn}}|^2, \label{eq:apprwa}\\
 \gamma &= \frac{A_{\rm He}A_{\rm Be}}{A}\nu_{BB}.\nonumber
\end{align}
where $\nu_{BB}$ denotes the width parameter of the Gaussian wave packet of Brink-Bloch wave function.

\begin{figure}[h]
 \centering
 \includegraphics[width=1.0\hsize]{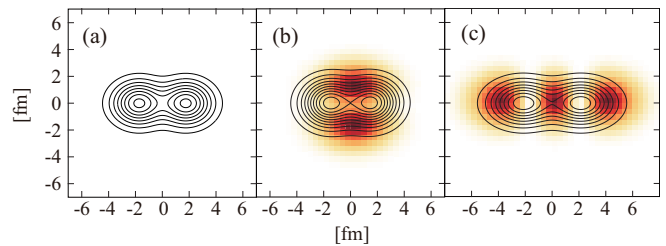}
 \caption{(color online) The density distribution of (a) $^{8}{\rm Be}$ and $^{10}{\rm Be}$
 with valence neutrons in (b) $\pi$-orbit and (c) $\sigma$-orbit.
 The contour lines show the proton density distributions.
 The color plots show the single particle orbits occupied by the most weakly bound neutron.} 
 \label{fig:bedensity}
\end{figure}

In this study, the Brink-Bloch wave function is constructed as follows.
First, the intrinsic wave function for $^{10}$Be and $^{8}$Be denoted by $\psi_{\rm Be}$ are generated by the AMD energy variation.
The intrinsic wave function of Be is approximated by a single AMD Slater determinant with spherical Gaussian wave packets with the width parameter $\nu_{BB}=0.16$ fm$^{-2}$.
In the case of $^8{\rm Be}$, the wave functions of the $0^+_1$ and $2^+_1$ states are calculated by the bound-state approximation.
The density distribution of obtained intrinsic wave function of $^8{\rm Be}$ is shown in Fig. \ref{fig:bedensity}(a) in which the distance between two $\alpha$ clusters is approximately 3.4 fm.
For $^{10}{\rm Be}$, we obtained two different intrinsic wave functions shown in Fig. \ref{fig:bedensity} (b) and (c) in which two valence neutrons occupy so-called $\pi$- and $\sigma$-orbits, respectively.
We regard that the former correspond to the ground band (the $0^+_1$, $2^+_1$, and $4^+_1$ states), while the latter is the excited band (the $0^+_2$, $2^+_3$, and $4^+_2$ states).
Then it is projected to the eigenstate of the angular momentum $j^\pi$ as $\phi_{\rm Be}(j^{\pi}_n)=\hat{P}^{j\pi}_{m0}\psi_{\rm Be}$, where we approximate that the Be wave function is axially symmetric The construction of the wave function of $^{4}$He and $^{6}$He is explained in the next section. 
The Brink-Bloch wave function is constructed by placing these He and Be wave functions with the inter-cluster distance $a$,
\begin{align}
 \Phi^{jm}_{BB}(a) &= {\mathcal A} 
 \biggl\{\phi_{\rm He}\left(-\frac{A_{\rm Be}}{A}a\right)
 \hat{P}^{j}_{m0}\psi_{\rm Be}\left(\frac{A_{\rm He}}{A}a\right)\biggl\}, \label{eq:bbwf} 
\end{align}
and projected to the eigenstate of the total spin-parity $J^\Pi$ as $\hat{P}^{J^\Pi}_{Mm}\Phi^{jm}_{BB}(a)$.
Then, we construct the wave function, in which the angular momentum $l$ of the inter-cluster motion and the angular momentum $j$ of Be are coupled to the total spin-parity $J^\Pi$, by summing up for all possible values of $m$,
\begin{align}
 \Phi^{J\pi jl}_{BB}(a) &=
 n\frac{2l+1}{2J+1}\sum_mC^{Jm}_{l0jm}\hat{P}^{J^\pi}_{Mm}\Phi^{jm}_{BB}(a),
 \label{eq:norm}
\end{align}
where $C^{Jm}_{l0jm}$ and $n$ denotes the Clebsch-Gordan coefficient and the normalization factor. 

Generally, the partial decay width should be independent on the choice of the channel radius.
However, in the practical calculation, the channel radius must be properly chosen to stabilize the results because of the following two problems. Firstly, the channel radius should not be too large value, because we adopt the bound-state approximation in the GCM calculation
and hence the wave function is not correct at large inter-cluster distance.
Secondly, the channel radius $a$ should not be too small, because the antisymmetrization effect cannot be neglected and the approximation is not valid.
Therefore, we adopted two different values  for the channel radius.
The first choice is $a=5.2$ fm which is common to the value used in the {\it R}-matrix analysis of the $\pi$-bond linear chain candidates observed in Ref.\cite{free14} and close to that in Ref.\cite{yama17}.
Unfortunately, this choice of channel radius was found inappropriate for the analysis of the $\sigma$-bond linear chain.
Because $\sigma$-bond linear chain is dominated by the $^{10}{\rm Be}(0^+_2,2^+_3)+\alpha$ channels and $^{10}{\rm Be}(0^+_2,2^+_3)$ have larger radii than the ground state, the larger channel radius should be used to avoid the antisymmerzation effect. Hence, we used $a=6.0$ fm for the analysis of $\sigma$-bond linear chain.

\subsection{$^{6}$He reduced width amplitude}
Here, we explain how the wave functions of $^{4}$He and $^{6}$He clusters are constructed.
The wave function of $^{4}$He is approximated by a $(0s_{1/2})^4$ wave function of harmonic oscillator (H.O.) which is represented by the Gaussian wave packet with the width of $\nu_{BB}=0.16$ fm$^{-2}$, 
\begin{align}
\phi_{0s}({\bm r})=(2\nu/\pi)^{3/4}e^{-\nu r^2}\otimes\chi.
\end{align}
The ground state of $^{6}$He is approximated by a $(0s_{1/2})^4(0p_{3/2})^2$ configuration as
\begin{align}
\phi_{\rm He}^{J=0} &= {\mathcal A}\{(0s_{1/2})^4[0p_{3/2}\otimes 0p_{3/2}]^{J=0}\} \label{eq:he6}
\end{align}
where $0p_{3/2}$ is also the eigen function of H.O.
In the practical calculation, we do not use H.O. wave functions directly, but the $0p_{3/2}$ wave function is represented by the sum of the infinitesimally shifted Gaussian wave packets $\phi_{0s}({\bm r}; {\bm \epsilon})=(2\nu/\pi)^{3/4}{\rm exp}{-\nu({\bm r}-{\bm \epsilon}^2})\otimes\chi$.
This greatly reduces the computational cost because it is possible to use ordinary computational code for AMD to calculate Eq. (\ref{eq:apprwa}).
The relationship between the shifted Gaussian wave packets and H.O. wave function is given as follows to the first order of the shift ${\bm \epsilon}$,
\begin{align}
\begin{split}
&\phi_{0s}({\bm r}; {\bm \epsilon})-\phi_{0s}({\bm r}; {\bm 0}) \\
&= \Bigl(\frac{2\nu}{\pi}\Bigr)^{3/4}\biggl\{ e^{-\nu({\bm r} - {\bm \epsilon})^2} - e^{-\nu r^2} \biggl\} \otimes (a\chi_\uparrow + b\chi_\downarrow) \\
 &\simeq \Bigl(\frac{2\nu}{\pi}\Bigr)^{3/4}2\nu {\bm r}\cdot{\bm \epsilon}e^{-\nu r^2}\otimes (a\chi_\uparrow + b\chi_\downarrow) \\
&= -\Bigl(\frac{2\nu}{\pi}\Bigr)^{3/4}2\nu re^{-\nu r^2}\biggl[a\mathcal{Y}_{1-1}({\bm \epsilon})
\phi_{p_{3/2\ 3/2}} \\
&\quad+ b\mathcal{Y}_{11}({\bm \epsilon})\phi_{p_{3/2\ -3/2}}\\
&\quad+\sqrt{\frac{1}{3}} \biggl\{ b\mathcal{Y}_{1-1}({\bm \epsilon})
-\sqrt{2}a\mathcal{Y}_{10}({\bm \epsilon}) \biggl\}\phi_{p_{3/2\ 1/2}}\\
&\quad+\sqrt{\frac{1}{3}} \biggl\{ a\mathcal{Y}_{11}({\bm \epsilon})
-\sqrt{2}b\mathcal{Y}_{10}({\bm \epsilon}) \biggl\}\phi_{p_{3/2\ -1/2}}\\
&\quad+\sqrt{\frac{1}{3}} \biggl\{ \sqrt{2}b\mathcal{Y}_{1-1}({\bm \epsilon})
+a\mathcal{Y}_{10}({\bm \epsilon}) \biggl\}\phi_{p_{1/2\ 1/2}}\\ 
&\quad-\sqrt{\frac{1}{3}} \biggl\{ \sqrt{2}a\mathcal{Y}_{11}({\bm \epsilon})
+b\mathcal{Y}_{10}({\bm \epsilon}) \biggl\}\phi_{p_{1/2\ -1/2}} \biggl] \label{eq:shiftg}
\end{split}
\end{align}
where $\mathcal{Y}_{1m}({\bm r})$ is the regular solid spherical harmonics,
\begin{align}
\mathcal{Y}_{1m}({\bm r}) &= \sqrt{\frac{4\pi}{3}}rY_{1m}(\hat{r}) \label{eq:harm},
\end{align}
where $\phi_{p_{3/2\ m}}$ and $\phi_{p_{1/2\ m}}$ denote the $0p_{3/2}$ and $0p_{1/2}$ wave functions with the magnetic quantum number $m$.
From Eq.(\ref{eq:shiftg}), we see that $0p_{3/2}$ wave functions can be described by the sum of the $\phi_{0s}({\bm r}; {\bm \epsilon})$ with proper choice of $a,b$ and ${\bm \epsilon}$.
Thus, Eq.(\ref{eq:he6}) is represented by the sum of the Slater determinant of the shifted Gaussian packets.
In the practical calculation the magnitude of ${\bm \epsilon}$ is chosen as $|{\bm \epsilon}|=0.02$.

\section{Results and Discussion}
In Sec. III A, we summarize the properties of the $\pi$-bond and $\sigma$-bond linear chains studied in our previous work \cite{baba16}.
In Sec. III B and C, by referring the latest experimental data and the theoretical analysis of the decay modes, we discuss the assignment of the linear-chain bands.

\subsection{Calculated linear-chain bands}

\begin{figure*}[t]
 \includegraphics[width=0.9\hsize]{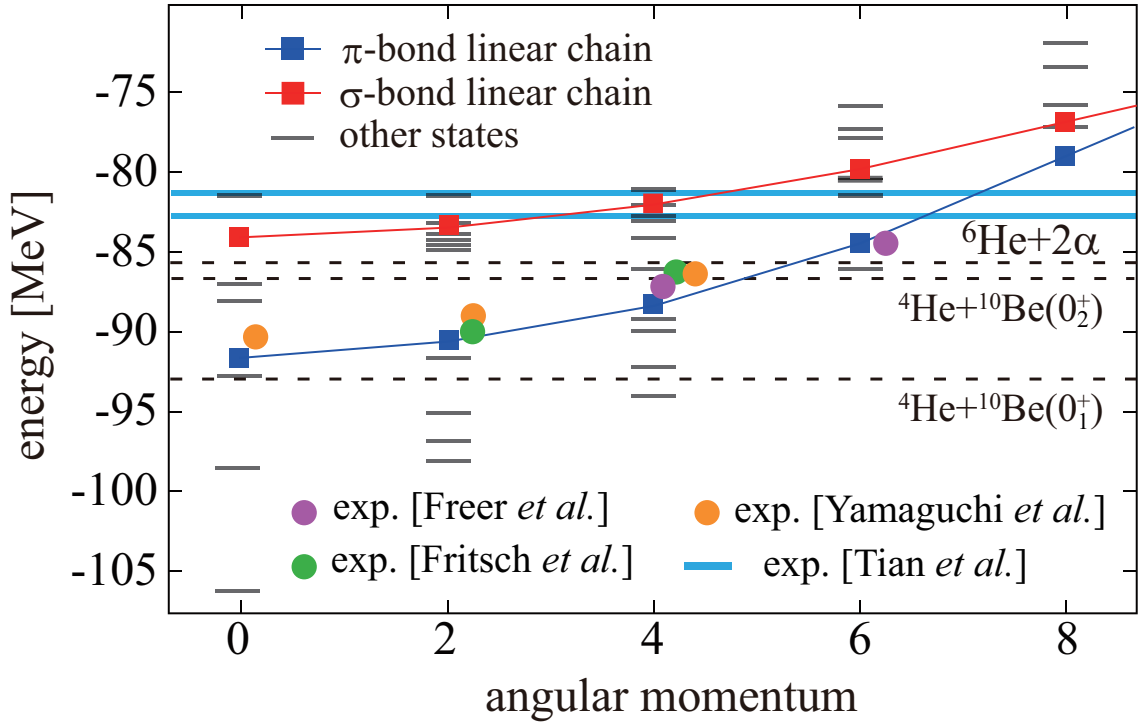}
 \caption{(color online) The positive-parity energy levels up  to $J^\pi=8^+$. 
 Filled circles show the resonances with the definite spin-parity assignments observed in the
 $\alpha+{}^{10}{\rm Be}$ resonant scattering\cite{free14,frit16,yama17}.
 Blue lines show the resonances without spin-parity assignment observed in the
 breakup reaction \cite{tian16}.
 Filled boxes show the calculated linear-chain bands, while lines show the other states.
 Only even angular momenta are shown.}
 \label{fig:spectrum+} 
\end{figure*}

\begin{figure}[h]
 \centering
 \includegraphics[width=1.0\hsize]{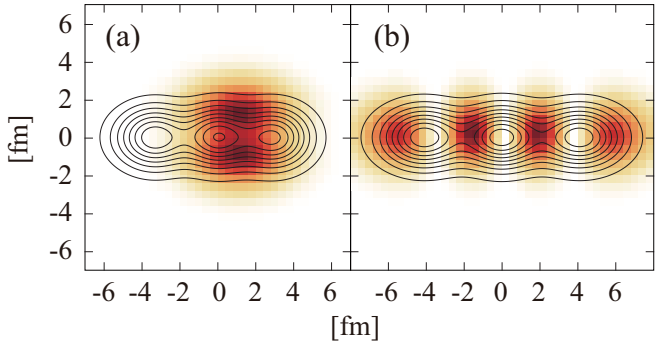}
 \caption{(color online) The density distribution of the linear-chain configurations of
 $^{14}{\rm C}$, with valence neutrons in (a) $\pi$-orbit and (b) $\sigma$-orbit.
 The contour lines show the proton density distributions.
 The color plots show the single particle orbits occupied by the most weakly bound neutron.} 
 \label{fig:density}
\end{figure}

\begin{figure}[h]
 \centering
 \includegraphics[width=1.0\hsize]{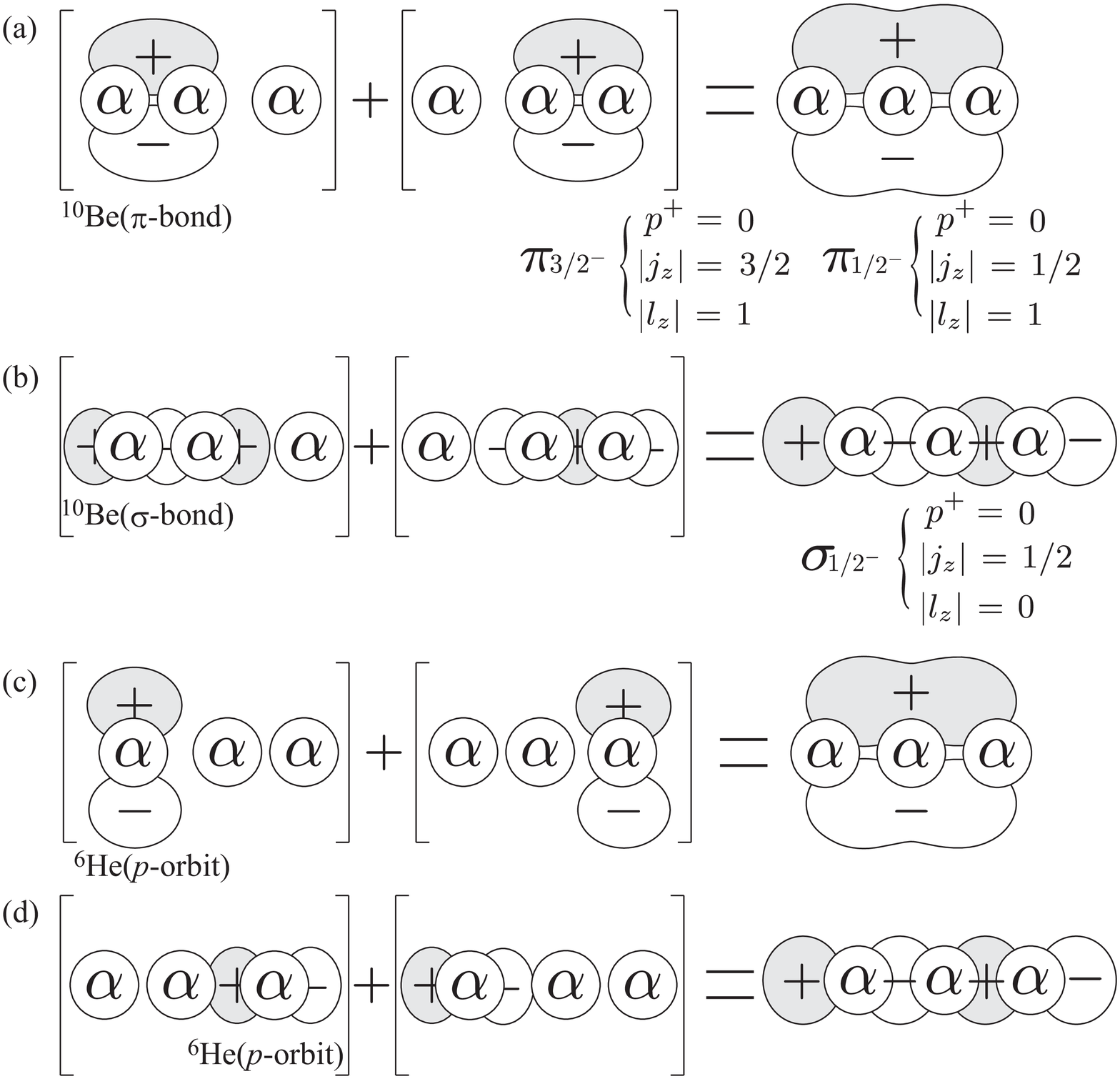}
 \caption{The schematic figure showing the $\pi$ and $\sigma$-orbits around the linear chain. 
 The combination of the $p$ orbits perpendicular to the symmetry axis generates
 $\pi$ orbits, while the combination of parallel orbits generates $\sigma$ orbit.} 
 \label{fig:illust_mol}
\end{figure}

Figure \ref{fig:spectrum+} summarizes the calculated rotational bands with the linear-chain configurations presented in Ref. \cite{baba16} and experimental data \cite{free14,frit16,yama17,tian16}.
The $\pi$-bond linear-chain band shown by blue squares is built on the $0^+$ state at 14.6 MeV which lies just above the $\alpha+{}^{10}{\rm Be}(0^+_1)$ threshold but below $\alpha+{}^{10}{\rm Be}(0_2^+)$ and $^{6}{\rm He}+{}^{8}{\rm Be}(0^+)$ thresholds.
The other band, the $\sigma$-bond linear-chain, is built on the $0^+$ state at 22.2 MeV which is above all of those thresholds.

Theoretically, the assignment of these two bands is rather unique.
The reason of the assignment and the properties of the linear-chain bands are as follows.
Firstly, these bands are dominated by the intrinsic states having the linear-chain configurations shown in Fig. \ref{fig:density}.
The $\pi$-bond linear chain has large overlap with the intrinsic wave function shown in Fig. \ref{fig:density}(a) which amounts to 87\% for the $0^+$ state at 14.6 MeV.
The proton density distribution shown by solid lines clearly indicates the formation of the linearly aligned three alpha clusters surrounded by the two valence neutrons shown by the color plot.
As already discussed in Refs.\cite{baba14,baba16}, in terms of the molecular orbit picture, this valence neutron orbit is interpreted as the $\pi$-orbit which is composed of the perpendicular alignment of the {\it p}-wave around the alpha cluster as illustrated in Fig.\ref{fig:illust_mol}(a).
The $\sigma$-bond linear-chain is dominated by the intrinsic wave function shown in Fig. \ref{fig:density}(b) whose overlap with the $0^+$ state at 22.2 MeV amounts to 99\%.
Again, we recognize the formation of the 3$\alpha$ linear chain, but the valence neutron orbit is different.
It is interpreted as the $\sigma$-orbit shown in Fig.\ref{fig:illust_mol}(b) which is composed of the parallel alignment of the {\it p}-orbits.
Since all other states in this energy region denoted by lines in Fig. \ref{fig:spectrum+} have much less overlap with the configurations, the linear-chain bands can be clearly assigned.

Secondly, the B(E2) transition strengths between the member states of these bands are rather strong compared to other in-band and inter-band transitions as listed in Table.III in Ref. \cite{baba16}, which is consistent with the dominance of the strongly deformed intrinsic shapes with linear-chain configuration.
Because the $\sigma$-bond linear chain is more strongly deformed than the $\pi$-bond linear chain, its in-band transition strengths are stronger than those of $\pi$-bond linear chain.
Deference of the deformation also reflected to the moment-of-inertia; $\hbar/2\Im=$179 and 98 keV for $\pi$-bond and $\sigma$-bond linear-chain bands, respectively.

Finally, among the excited states located around the $\alpha+{}^{10}{\rm Be}(0^+_1)$, $\alpha+{}^{10}{\rm Be}(0^+_2)$ and $^{6}{\rm He}+{}^{8}{\rm Be}(0^+)$ thresholds, the linear-chain bands have largest $\alpha$ and $^{6}{\rm He}$ reduced widths.
Therefore, the alpha decaying resonances in the vicinity of these thresholds are, if observed, regarded as the candidates of the linear-chain bands.

\subsection{Resonances observed in the $\alpha+{}^{10}{\rm Be}$ scattering and the assignment of the $\pi$-bond linear chain band}
Now, we discuss the assignment of the linear-chain bands based on the latest experimental data.
Freer {\it et al.} \cite{free14}, Fritsch {\it et al.} \cite{frit16} and more recently, Yamaguchi {\it et al.} \cite{yama17} independently reported the resonances observed in the $\alpha+{}^{10}{\rm Be}$ scattering, which are shown by circles in Fig. \ref{fig:spectrum+} and summarized in Table. \ref{table:lc}.
Freer {\it et al.} reported the $4^+$ and $6^+$ resonances at 18.22 and 20.80 MeV, respectively, while Fritsch {\it et al.} reported the $2^+$ and $4^+$ resonances at 15.0 and 19.0 MeV.
A candidate of the $2^+$ resonance at 17.95 MeV was also reported by Freer {\it et al.}, but not shown in Fig. \ref{fig:spectrum+} because the spin-parity assignment is not so firm.
Yamaguchi {\it et al.} reported the $0^+$, $2^+$ and $4^+$ resonances at 15.07, 16.22 and 18.87 MeV.
The $4^+$ energy is very close to that observed by Fritsch {\it et al.} and the $0^+$ state may correspond to the Fritsch {\it et al.}'s 15.0 MeV state which was assigned as $2^+$.

Although they suggest different spin-parity assignments, we consider that they observed essentially the same resonances which are assigned to the $\pi$-bond linear-chain band from the following reasons.
Firstly, it is clear that these resonance energies very nicely agree with those of the calculated $\pi$-bond linear-chain, regardless of the spin-parity assignments.
Furthermore, the observed data show the large moment-of-inertia of the band; $\hbar/2\Im=$ 116 keV \cite{free14} and 190 keV \cite{yama17}.
In any cases, the very large moment-of-inertia are consistent with the large deformation of the linear-chain band which reaches to 3:1 ratio of the deformation axes.
In particular, the moment-of-inertia reported by Yamaguchi {\it et al.} ($\hbar/2\Im=$190 keV) is very close to the present result.
Since their experiments have better statistics and larger angular coverage than others, we expect that their spin-parity assignments are reliable.
We also note that the 15.0 MeV state observed by Fritsch {\it et al.} can be assigned as $0^+$ instead of $2^+$, because this state is very close to the $0^+$ state at 15.07 MeV reported by Yamaguchi {\it et al.}
With this change of the assignment, the moment-of-inertia of Fritsch {\it et al.}'s experiment is close to the Yamaguchi {\it et al.}'s data and consistent with the present theoretical result.

Secondly, as listed in Table. \ref{table:lc}, the observed resonances have large alpha decay widths to the $\alpha+{}^{10}{\rm Be}(0^+_1)$ channel comparable with those of the $\pi$-bond linear-chain.
Experimental data are not quantitatively consistent to each other, but most of them are few hundreds keV which are the same order of magnitude with the calculated $\pi$-bond linear-chain.
This is rather strong evidence of the linear-chain formation, because theories predict no other states than the $\pi$-bond linear-chain states which have large alpha decay widths in this energy region.
It must be noted that the $\sigma$-bond linear-chain band has rather small decay widths to he $\alpha+{}^{10}{\rm Be}(0^+_1)$ channel, which distinguishes the $\sigma$-bond linear-chain from the $\pi$-bond linear-chain.
The reason for this decay suppression will be explained in the next section.

Finally, theories predicted the decay of the $\pi$-bond linear-chain to the $\alpha+{}^{10}{\rm Be}(2^+_1)$ channel despite of the smaller decay {\it Q}-value (Table.\ref{table:lc2}).
This is because of the strong admixture of the $\alpha+{}^{10}{\rm Be}(0^+_1)$ and $\alpha+{}^{10}{\rm Be}(2^+_1)$ configurations in the $\pi$-bond linear-chain, which originates in the strong coupling nature of the linearly aligned alpha clusters.
Experimentally, the width of the $\alpha+{}^{10}{\rm Be}(2^+_1)$ decay has not been measured, but Fritsch {\it et al.} reported the decay of the $4^+$ resonance to the $\alpha+{}^{10}{\rm Be}(2^+_1)$ channel.
Thus, the excitation energies, moment-of-inertia and the decay widths are consistent between the theory and the $\alpha+{}^{10}{\rm Be}$ scattering experiment, and hence, the formation of $\pi$-bond linear-chain in $^{14}$C looks rather plausible.
We also note that the same resonances were also observed in the break up \cite{haig08} and transfer reactions \cite{tian16,li17}, although the spin-parity assignment was not given.

\begin{table*}[h]
 \caption{Excitation energies (MeV) and  $\alpha$-decay widths (keV) up to $J^\pi=8^+$ of the linear-chain states and the experiments. $\Gamma_\alpha$ shows the decay to the ground state ($0^+_1$) of $^{10}{\rm Be}$. Numbers in parentheses are tentative.}  
\label{table:lc}
\begin{center}
 \begin{ruledtabular}
  \begin{tabular}{lccccccccccc}
  \multicolumn{1}{}{}
  & \multicolumn{3}{c}{$\pi$-bond linear chain}
  & \multicolumn{2}{c}{$\sigma$-bond linear chain}
  & \multicolumn{2}{c}{exp. \cite{free14}}
  & \multicolumn{2}{c}{exp. \cite{frit16}}
  & \multicolumn{2}{c}{exp. \cite{yama17}} \\ 
   $J^\pi$ & $E_x$ & $\Gamma_\alpha(5.2 {\rm fm})$ & $\Gamma_\alpha(6.0 {\rm fm})$ & $E_x$ & $\Gamma_\alpha(6.0 {\rm fm})$ & $E_x$ & $\Gamma_\alpha$ & $E_x$ & $\Gamma_\alpha$ & $E_x$ & $\Gamma_\alpha$ \\
   \hline
   $0^+$ & 14.64 & 250 & 179 & 22.16 & 0.2 & & & & & 15.07 & 760 \\
   $2^+$ & 15.73 & 214 & 188 & 22.93 & 0.4 & (17.95) & (760) & 15.0 & 290 & 16.22 & 190 \\
   $4^+$ & 17.98 & 149 & 147 & 24.30 & 0.3 & 18.22 & 200 & 19.0 & 340 & 18.87 & 45 \\
   $6^+$ & 21.80 & 123 & 151 & 26.45 & 0.2 & 20.80 & 300 & & & & \\
   $8^+$ & 27.25 & 77 & 120 & 29.39 & 0.2 & & & & & & \\
  \end{tabular}
 \end{ruledtabular}
 \end{center}
\end{table*}

\begin{table*}[h]
 \caption{Excitation energies (MeV) and  $\alpha$-decay widths (keV) 
 to the $2^+_1$ state of $^{10}{\rm Be}$.}  
\label{table:lc2}
\begin{center}
 \begin{ruledtabular}
  \begin{tabular}{lccccc}
  \multicolumn{1}{}{}
  & \multicolumn{3}{c}{$\pi$-bond linear-chain}
  & \multicolumn{2}{c}{$\sigma$-bond linear-chain} \\ 
   $J^\pi$ & $E_x$ & $\Gamma_\alpha(5.2 {\rm fm})$ & $\Gamma_\alpha(6.0 {\rm fm})$ & $E_x$ & $\Gamma_\alpha(6.0 {\rm fm})$ \\
   \hline
   $0^+$ & 14.64 & - & - & 22.16 & 0.6 \\
   $2^+$ & 15.73 & - & - & 22.93 & 0.2 \\
   $4^+$ & 17.98 & 118 & 111 & 24.30 & 1.8 \\
   $6^+$ & 21.80 & 256 & 271 & 26.45 & 0.4 \\
   $8^+$ & 27.25 & 397 & 421 & 29.39 & 0.8 \\
  \end{tabular}
 \end{ruledtabular}
 \end{center}
\end{table*}

\subsection{Higher-lying resonances observed in the break-up reaction and the assignment of the $\sigma$-bond linear-chain band}

\begin{figure*}[h]
 \centering
 \includegraphics[width=0.9\hsize]{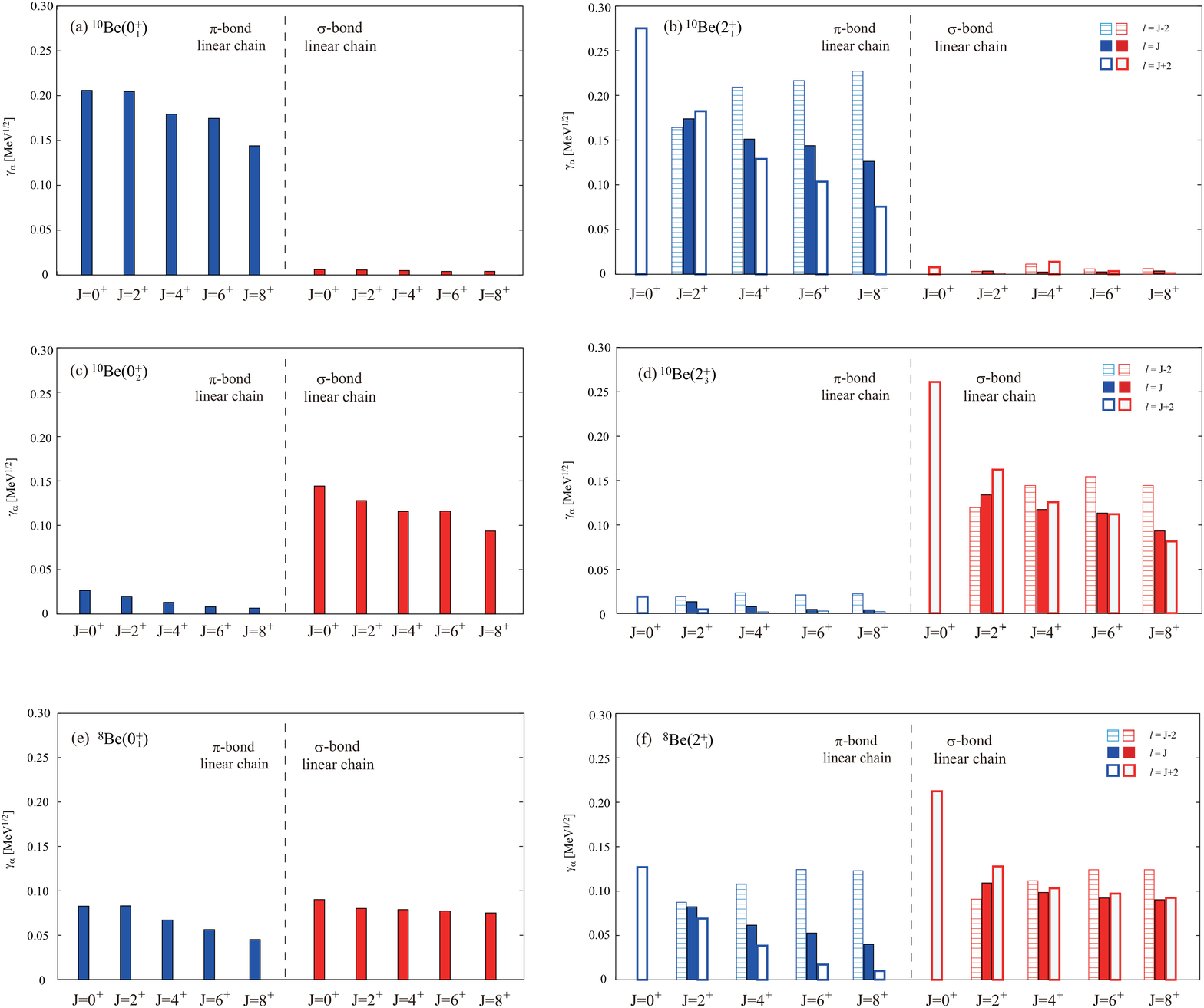}
 \caption{(color online) The calculated $\alpha$- and $^{6}{\rm He}$-decay reduced widths up to $J^\pi=8^+$.
 Panels (a)(b) show the decay to the ground band of $^{10}{\rm Be}$ ($\pi$-bonded $^{10}{\rm Be}$). 
 Panels (c)(d) show the decay to the excited band of $^{10}{\rm Be}$ ($\sigma$-bonded $^{10}{\rm Be}$). 
 Panels (e)(f) show the decay to the ground band of $^{8}{\rm Be}$.
 The $l$ denotes the relative motion between ${\rm Be}(2^+)$ and $\alpha$ particle.
 The channel radius $a$ is 6.0 fm.}  
 \label{fig:rwa+} 
\end{figure*}

Quite recently, in addition to the candidates of the $\pi$-bond linear chain, Tian {\it et al.} \cite{tian16} and Li {\it et al.} \cite{li17} reported new resonances at 22.4 and 24.0 MeV observed in the $^{9}{\rm Be}(^{9}{\rm Be}, \alpha+{}^{10}{\rm Be})\alpha$ reaction.
Since their spin-parity were not assigned yet, they are shown by blue lines in Fig. \ref{fig:spectrum+}.
We see that their energies are very close to those of the calculated $2^+$, $4^+$ and $6^+$ states of the $\sigma$-bond linear chain, but we cannot exclude the assignment to the $6^+$ or $8^+$ states of the $\pi$-bond linear chain.

In order to identify the structure of these resonances,
we focus on the decay patterns of the $\pi$- and $\sigma$-bond linear chains.
The reduced widths to various decay channels summarized in Figure \ref{fig:rwa+} suggests unique decay patterns of the linear chains.
From the panels (a) to (d), we see that all of the $\pi$-bond linear chain states decay to the ground band of $^{10}$Be ($0^+_1$ and $2^+_1$), but not to the excited band ($0^+_2$ and $2^+_3$).
On the other hand, the $\sigma$-bond linear chain has quite the opposite pattern; it decays to the excited band, but not to the ground band.
This clearly distinguishes two linear-chains, and the reason of the difference is qualitatively understood from the intrinsic density distributions of the $^{10}$Be and linear-chains shown in Figs.\ref{fig:bedensity} and \ref{fig:density}.
Both of the ground band of $^{10}$Be and $\pi$-bond linear chain has $\pi$-bonding neutrons,
and hence, the $\pi$-bond linear chain can be described by the linear alignment of the $^{10}$Be($0^+_1$ and $2^+_1$) and alpha cluster as illustrated in Fig.\ref{fig:illust_mol} (a).
Since this configuration is orthogonal to the $\alpha+^{10}$Be($0^+_1$ and $2^+_1$) the decay suppression to the $\alpha+^{10}$Be($0^+_1$ and $2^+_1$) channels can be naturally understood.
In the same way, the $\sigma$-bond linear chain can be described by the linear alignment of the 
$^{10}$Be($0^+_2$ and $2^+_3$)$+\alpha$ as shown in Fig.\ref{fig:illust_mol} (b) which explains the decay pattern of the $\sigma$-bond linear chain.

Experimentally, Li {\it et al.} \cite{li17} reported that the resonances at 22.4 and 24.0 MeV dominantly decay to the 6 MeV state of $^{10}{\rm Be}$ which is deduced to be the $0^+_2$ state of $^{10}{\rm Be}$.
Therefore, we conclude that these new resonances are promising candidates of the $\sigma$-bond linear-chain.
Different from the calculated $\sigma$-bond linear-chain, it is reported that observed resonances also decay to the ground band of $^{10}{\rm Be}$.
This discrepancy may be explained as follows.
In the present calculation, we approximated that the ground and excited bands of $^{10}{\rm Be}$ have pure $\pi$- and $\sigma$-bond configurations, respectively.
However, in reality, it is known that there are admixture of these configurations and the $0^+_1$ and  $2^+_1$ states have non-negligible amount of the $\sigma$-bond configuration.
Therefore, it is natural that the observed resonances also decay to the ground band as well as the excited band of $^{10}{\rm Be}$.

The panels (e) and (f) show that both of the $\pi$- and $\sigma$-bond linear chains have large reduced widths in the $^{6}{\rm He}+{}^{8}{\rm Be}$ channel, which is another interesting feature of the linear-chains.
This is again schematically understood from Fig.\ref{fig:illust_mol}.
Because the valence neutron in $\pi$- and $\sigma$-orbits are covalent, the linear-chains can also be described by the linear alignment of the $^{6}{\rm He}+{}^{8}{\rm Be}$ as illustrated in Fig.\ref{fig:illust_mol} (c) and (d).
Therefore, the $\sigma$-bond linear chain and high-spin states ($J^\Pi \geq 6^+$) of $\pi$-bond linear chain which locate above the $^{6}{\rm He}+{}^{8}{\rm Be}$ threshold should also decay to the three-body final state through the sequential two body decays, $^{14}{\rm C}^*\rightarrow{}^6{\rm He}+{}^8{\rm Be}\rightarrow{}^6{\rm He}+\alpha+\alpha$.
As listed in Table.\ref{table:sigma}, the decay widths of the $\sigma$-bond linear chain to the $^{6}{\rm He}+{}^{8}{\rm Be}$ channel are in the same order with those in the $\alpha+{}^{10}{\rm Be}$ channel, and hence, the decay to $^{6}{\rm He}+\alpha+\alpha$ is another evidence of the linear-chain formation.

In conclusion, the $\pi$-bond linear-chain states decay to $^{10}{\rm Be}(\pi^2)+\alpha$ and higher-spin than the $6^+$ states can decay to $^{6}{\rm He}+\alpha+\alpha$.
In the case of the $\sigma$-bond linear-chain, they decay to $^{10}{\rm Be}(\sigma^2)+\alpha$ and $^{6}{\rm He}+\alpha+\alpha$ because all member states are above the both threshold energies.
This decay pattern is an important evidence to show the formation of two linear-chains in $^{14}$C.

\begin{table*}[h]
 \caption{Decay widths of three different channels (keV) up to $J^\pi=8^+$ of the $\sigma$-bond linear-chain states. The channel radius $a$ is 6.0 fm.}  
\label{table:sigma}
\begin{center}
 \begin{ruledtabular}
  \begin{tabular}{lcccc} 
   $J^\pi$ & $E_x$ & $\Gamma(^{4}{\rm He}+{}^{10}{\rm Be}(0^+_1;\pi^2))$ & $\Gamma(^{4}{\rm He}+{}^{10}{\rm Be}(0^+_2;\sigma^2))$ & $\Gamma(^{6}{\rm He}+{}^{8}{\rm Be})$ \\
   \hline
   $0^+$ & 22.16 & 0.2 & 136 & 38 \\
   $2^+$ & 22.93 & 0.4 & 99 & 29 \\
   $4^+$ & 24.30 & 0.3 & 63 & 23 \\
   $6^+$ & 26.45 & 0.2 & 42 & 17 \\
   $8^+$ & 29.39 & 0.2 & 17 & 13 \\
  \end{tabular}
 \end{ruledtabular}
 \end{center}
\end{table*}

\section{SUMMARY}
In this paper, we focus on the linear-chain states of $^{14}$C based on the AMD calculations
to establish the existence of the linear-chain configuration.  

The linear-chain configurations generate two rotational bands.
At strong deformed prolate region, two different linear-chain configurations with valence neutrons
in $\pi$-orbit and $\sigma$-orbit were obtained. 
The $\pi$-bond linear chain
generates a rotational band around the $\alpha$ threshold energy.
The energies and $\alpha$ decay widths of the $\pi$-bond linear chain are in reasonable agreement with the resonances observed by the $\alpha+{}^{10}{\rm Be}$.
Thus, the $\pi$-bond linear-chain formation in $^{14}{\rm C}$ looks plausible.

On the other hand, the $\sigma$-bond linear-chain generates a rotational band around the $^6{\rm He}$ threshold energy which is 7.5 MeV higher than the $\alpha$ threshold energy.
Newly observed resonance states are close to energies of both the low-spin states of the $\sigma$-bond linear-chain and the $6^+$ state of the $\pi$-bond linear-chain.
In order to distinguish the $\pi$- and $\sigma$-bond linear-chain,
we focus on the decay patterns of them.
Reduced widths show that the $\pi$-bond linear-chain states decay into the ground band of 
$^{10}{\rm Be}$, while the $\sigma$-bond linear-chain states decay into the excited band of
$^{10}{\rm Be}$.
This difference is due to the molecular-orbit of $^{10}{\rm Be}$.

From $^{6}{\rm He}$ reduced width, in addition, 
it is found that the $\sigma$-bond linear-chain states decay into not only the excited band of
$^{10}{\rm Be}$ but also $^{6}{\rm He} + \alpha + \alpha$. 
Furthermore, the calculation predicts that the linear-chain will also decay to the $^{8, 10}{\rm Be}(2^+)$ as well as to the ground state of $^{8, 10}{\rm Be}$.
This characteristic decay patterns are, if it is observed, a strong 
signature of the $\pi$- and $\sigma$-bond linear-chain formations.

\acknowledgements
The authors acknowledges the fruitful discussions with Dr. Suhara, Dr. Kanada-En'yo, Dr. Li, and Dr. Ye.
One of the authors (M.K.) acknowledges the support by the Grants-in-Aid for Scientific Research on Innovative Areas from MEXT (Grant No. 2404:24105008) and JSPS KAKENHI Grant No. 16K05339. 
The other author (T.B.) acknowledges the support by JSPS KAKENHI Grant No. 16J04889.

\end{document}